\documentclass[aps,twocolumn,prl]{revtex4-1} 
\usepackage{xcolor,amssymb}
\usepackage{graphicx}
\usepackage{amsmath,amssymb,bm,epstopdf}
\usepackage{graphicx,epsfig,epstopdf,dsfont,color}\usepackage{soul,dcolumn,chngcntr,,tikz}
\usepackage{graphicx,epsfig,epstopdf,dsfont,color}\usepackage{soul,dcolumn,chngcntr,,tikz}
\usepackage{amsmath,amssymb,caption}
\usepackage{bm,gensymb,floatrow,subfig}
\usepackage{multirow}
\usepackage{float}

\def\u{\bm{u}}
\def\ba#1\ea{\begin{align}#1\end{align}}
\def\bsa#1#2\esa{\begin{subequations}\label{#1}
\begin{align}#2\end{align} \end{subequations}}
\def\lp{\left(}
\def\rp{\right)}
\def\lb{\left[}
\def\rb{\right]}
\def\lcb{\left\{}

\def\NA{\bm{\nabla}}
\def\DEL{\nabla^2}
\def\f{\frac}
\def\b{\mathbf}
\def\p{\partial}

\def\NA{\nabla}

\begin{document}

\title{Order out of chaos: slowly-reversing mean flows emerge from \\   turbulently-generated internal waves}

\author{Louis-Alexandre Couston$^{1,*}$, Daniel Lecoanet$^{2}$, Benjamin Favier$^1$, Michael Le Bars$^1$}

\affiliation{$^1$ CNRS, Aix Marseille Univ, Centrale Marseille, IRPHE, Marseille, France \\ $^2$ Princeton Center for Theoretical Science, Princeton, NJ 08544, USA}

\begin{abstract}

We demonstrate via direct numerical simulations that a periodic, oscillating mean flow spontaneously develops from turbulently-generated internal waves. We consider a minimal physical model where the fluid self-organizes in a convective layer adjacent to a stably-stratified one. Internal waves are excited by turbulent convective motions, then non-linearly interact to produce a mean flow reversing on time scales much longer than the waves' period. Our results demonstrate for the first time that the three-scale dynamics due to convection, waves, and mean flow, is generic and hence can occur in many astro/geophysical fluids. We discuss efforts to reproduce the mean flow in reduced models, where the turbulence is bypassed. We demonstrate that wave intermittency, resulting from the chaotic nature of convection, plays a key role in the mean-flow dynamics, which thus cannot be captured using only second-order statistics of the turbulent motions.

\end{abstract}






\maketitle


An outstanding question in fluid dynamics is whether large-scale flows can be accurately captured in reduced models that do not resolve fluid motions on small spatio-temporal scales. Reduced models are necessary in many fields of fluid mechanics, since fluid phenomena often occur on a wide range of spatial and temporal scales, preventing exploration via direct numerical simulations (DNS) of the Navier-Stokes equations. This question is of interest to, for instance, the turbulence community, which has developed closure models in Large-Eddy Simulations and Reynolds-Averaged Navier-Stokes simulations \cite{durbin2018}; the statistical physics and geophysics communities, who aim to describe the self-organization and large-scale behavior of turbulent flows \cite{bouchet2012,Marston2016,Fauve2017,Pouquet2013}; atmospheric and oceanographic scientists, whose goals are to provide long-time predictions of the evolution of our climate using weather-ocean models with coarse resolution \cite{Bauer2015,Buijsman2016}. 

A drastic approximation would be to assume that large-scale flows and small-scale motions are dynamically decoupled, but this is rarely the case. A number of important slow large-scale flows are controlled by rapid processes at the small scales. For instance, the 22-year cycle of solar magnetism is driven by the Sun's convective interior, which evolves on month-long or shorter timescales \cite{Hathaway2010,Mininni2002}; upwelling of the planetary-scale thermohaline circulation of Earth's oceans hinges on enhanced mixing events that critically depend on small-scale ($\sim 100$ metres) internal waves \cite{Polzin1997,Nikurashin2013}; Jupiter's zonal jets develop from small-scale turbulence patterns due to convective heat transfers in the weather layer and deep interior \cite{Cabanes2017}. 

The generation of a large-scale flow by turbulent fluctuations can be studied by spatial-averaging the Navier-Stokes equations. Let us consider the case of a large-scale mean flow $\bar{u}$ in the horizontal $x$ direction perpendicular to downward gravity. We write $(u',w')$ the velocity fluctuations in $(x,z)$ directions with $\hat{z}$ the upward vertical axis. In these two dimensions, the horizontal-mean of the Navier--Stokes equation  in the $x$ direction reads
\ba{}\label{eq1}
\p_t \bar{u} - \nu\p_{zz}\bar{u} =  -\p_z \overline{(w'u')}, 
\ea
with $\nu$ the kinematic fluid viscosity. The right-hand side of \eqref{eq1} is minus the divergence of the Reynolds stress and is the momentum source or sink for the mean flow. In isotropic homogeneous turbulence, we do not expect the generation of a mean flow due to the lack of symmetry breaking. However, any inhomogeneity or anisotropy of the fluctuations can initiate a slowly-varying mean flow, whose fate depends on its interaction with the fluctuations \cite{Fauve2017}. The parameterization of the Reynolds stress $\overline{(w'u')}$ for unresolved scales is the key ingredient in all reduced models. Generally, a closure model expresses the Reynolds stresses in terms of the resolved variables \cite{durbin2018}.

In our case of interest, the small-scale fluctuations are oscillating disturbances of the density field called internal waves. Internal waves are ubiquituous in oceans \cite{Garrett1979}, planetary atmospheres \cite{Fritts2003,Miller2015,tellmann2013,piccialli2014}, stars \cite{Charbonnel2007,Straus2008}, brown dwarves \cite{Showman2013} and planetary cores \cite{Buffett2014}. In the atmosphere, internal waves actively contribute to the generation of mean equatorial winds in Earth's  stratosphere, which change direction roughly every 14 months, coined the Quasi-Biennial Oscillation (QBO) \cite{Baldwin2001}. Internal waves may also be involved in the generation of reversing zonal flows on Saturn \cite{Orton2008} and Jupiter \cite{leovy1991}, they are of interest for extrasolar planetary atmospheres \cite{Watkins2010}, and may influence the differential rotation of stars \cite{Rogers2012} and slow large-scale motions of Earth's magnetic field \cite{Livermore2017}.

\begin{figure*}[ht]
\centering
\includegraphics[width=17. cm]{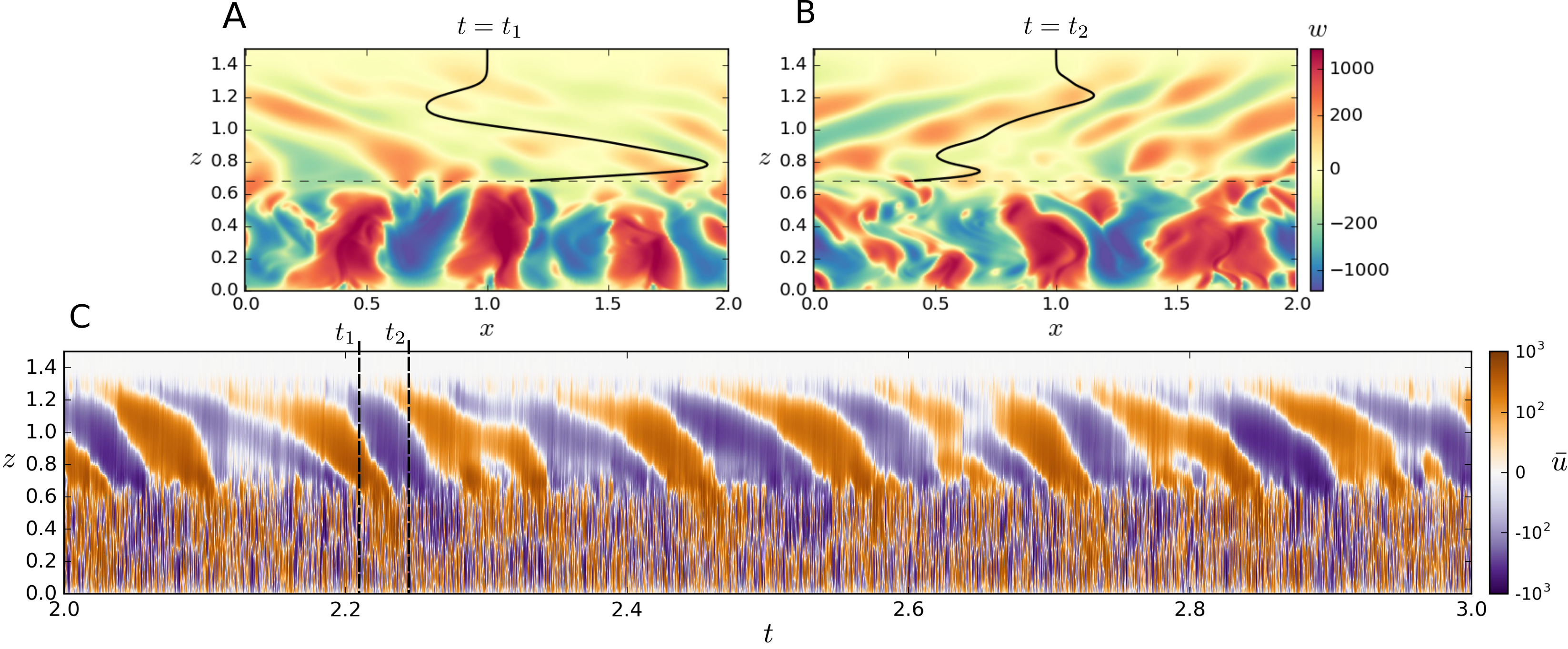}
\vspace{-0.1in}\caption{DNS results. (A) and (B) show snapshots of the vertical velocity field $w$ at times $t_1=2.21$ and $t_2=2.24$, along with the mean flow $\bar{u}$ (solid line) for $z>z_{NB}=0.68$ (with $\bar{u}=0$ corresponding to $x=1$). Vertical velocity patterns show convective motions in the lower part of the domain ($z\leq z_{NB}$) and internal-wave motions in the upper part ($z\geq z_{NB}$).  Note that energy propagates upward along wave crests, so crests toward the upper left (right) correspond to retrograde (prograde) waves. (C) shows the mean flow $\bar{u}(t,z)$. The mean flow in the convective zone corresponds to the average of stochastic plumes emitted from the bottom boundary, hence reverses on a relatively rapid, convective time scale. In the stably-stratified layer, $\bar{u}$ results from the nonlinear interaction of internal waves and oscillates on time scales $\sim 0.1$, much longer than the buoyancy period $\sim \pi 10^{-4}$. Simulation details and movies are available in Supplemental Material \cite{SM}.}\label{fig0}
\end{figure*}

Here, we report results of the first DNS of a realistic slowly-reversing mean flow in two dimensions, and we unravel the key physics of the generation mechanism using a hierarchy of low-order models in which the Reynolds stresses are approximated.  We use the horizontally-periodic self-consistent model of convective--stably-stratified dynamics of \cite{Couston2017}. The velocity $\b{u}=(u,w)$, temperature $T$, and density anomaly $\rho=-\alpha T$ satisfy the Boussinesq equations 
\bsa{a}
& \p_t \b{u} + (\u\cdot\NA)\b{u} + \NA  p = Pr\DEL \b{u} - Pr Ra  \rho\hat{z} - \b{u}\tau, \\ 
&\p_t T + (\u\cdot\NA) T = \DEL T, \\ \label{a4}
&\NA\cdot \u = 0, 
\esa
non-dimensionalized with $\kappa$ (thermal diffusivity) and  $H$ (characteristic height). The fluid is thermally stratified ($T_t$ and $T_b$ imposed on the top and bottom no-slip boundaries) and exhibits a buoyancy reversal at the inversion temperature $T_i$ with $T_b>T_i>T_t$ (similar to water whose density maximum is at $4^{\circ}C$ \cite{Lebars2015}). Thus the fluid spontaneously organizes into a lower, nearly isothermal convective region, and an upper stably stratified region. $Pr=\nu/\kappa$ and $Ra=\alpha_s g\Delta TH^3/(\kappa\nu)$ are the Prandtl and global Rayleigh numbers; $\alpha_s$ is the expansion coefficient for $T>T_i$; and $\Delta T>0$ is the difference between the dimensional bottom and inversion temperatures, such that using $T_i=0$ as the dimensionless reference temperature, we have $T_b=1$. The buoyancy reversal is obtained using the nonlinear equation of state for $\rho$:
\ba{}\label{b}
\rho(T)=-\alpha(T) T=\lcb \begin{array}{c}
-T,~T\geq T_i=0, \\
ST ,~T\leq T_i=0,
\end{array}\right.
\ea
with $S$ the stiffness parameter \cite{Couston2017}. We define the neutral buoyancy level $z_{NB}$ to be the height where adiabatic plumes emitted from the bottom boundary become neutrally buoyant.  This corresponds to the height of the convection zone \cite{Lecoanet2016,Couston2017}, or equivalently, the base of the stable layer (dashed lines in figures \ref{fig0}A-B). The normalized domain lengths are $L_x=2$, $L_z=1.5$ in the $x$, $z$ directions, which leads to an aspect ratio of the convection at statistical steady-state close to 3 for all simulations; $\tau=10^2\sqrt{2}\{\tanh[ (z-L_z+0.15)/0.05 ]+1\}/2$ is a $z$-dependent linear damping used to prevent wave reflections from the top boundary. We solve equations \eqref{a} via DNS using Dedalus \cite{Burns2017} with Chebyshev polynomials (Fourier modes) in $z$ ($x$) direction. DNS are run over several thermal diffusion times in order to allow the system to reach a statistical equilibrium self consistently, and obtain several reversals of the mean flow.

Figure \ref{fig0} shows the main DNS results of the paper, obtained for $T_t=-43$, $T_b=1$, $Pr=0.2$, $Ra=1.2\times 10^8$ and $S=1/3$, such that the convection-wave coupling is relatively strong and the interface is flexible \cite{Couston2017}. With $z_{NB}=0.68$, the effective Rayleigh number is $Ra_{\rm eff}=z_{NB}^3Ra\approx 4 \times 10^7$. Snapshots of vertical velocity (figures \ref{fig0}A,B) reveal large convective updrafts and downflows below $z_{NB}$, and internal waves above.  If there was no mean flow in the stably stratified layer, convection would generate prograde and retrograde waves with similar amplitude. However, in figures \ref{fig0}A,B, the internal waves are mostly propagating in a single direction, an indication that the mean-flow (shown by the solid line) is filtering waves going in the opposite direction. The evolution of the mean flow over one thermal time scale is shown in figure \ref{fig0}C. The stable layer has a strong mean-flow which reverses every $\sim 0.05$ thermal time. Each new mean-flow phase starts near the top of the domain and descends toward the convective layer. The mean flow is driven by wave damping at critical layers and by viscous and thermal dissipation. Critical layers are ubiquitous in our DNS because convection generates a broad spectrum of waves, some of which have low phase velocities. Viscous and thermal dissipation effects are relatively strong in our DNS, so the basic mean-flow mechanism is essentially due to wave dissipation.

Previous studies of wave--mean-flow interactions have focused on momentum-deposition by internal waves of a single frequency and wavenumber \cite{Plumb1978,Semin2016}. In such cases, it can be shown analytically that a slowly-reversing mean flow emerges provided that there are both prograde and retrograde waves, as well as an initial disturbance. The prograde (resp. retrograde) wave provides a $+x$ positive (resp. negative) acceleration for the mean-flow through damping. Then, the competition of the two forces (whose intensity depends on the direction of the mean flow through the Doppler shift) leads to the observed long-time oscillation of $\bar{u}$ \cite[][]{Baldwin2001}.

\begin{figure}
\centering
\includegraphics[width=.95\linewidth]{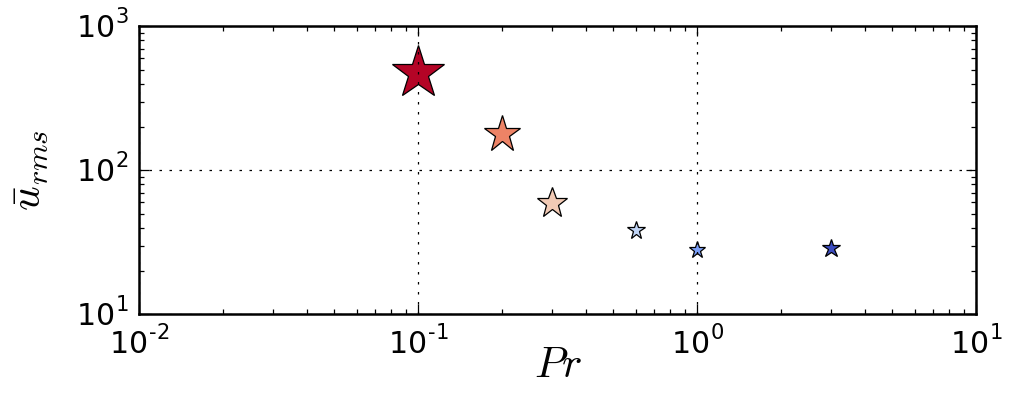}
\vspace{-0.1in}\caption{Mean flow rms $\bar{u}_{rms}$ as a function of $Pr$. The mean flow becomes stronger as $Pr$ decreases and is also more regular: the symbols' area is inversely proportional to the frequency bandwidth of $\bar{u}$, defined as the difference $\Delta f=f_{.9}-f_{.1}$ of the two frequencies $f_{.1}$ and $f_{.9}$ below and above which lies 10\% of the mean-flow energy.}
\label{figPr}
\end{figure}

Our results demonstrate for the first time that an oscillating mean flow can emerge from internal waves generated by turbulent motions with no control over the waves (i.e. no parameterization). Importantly, the fundamental mechanism that applies for monochromatic waves also applies for a broadband spectrum of internal waves: damping and momentum deposition is stronger for waves going in the same direction as the mean flow. This is shown in figure \ref{fig0}A where a strong mean flow in the positive direction  strongly dissipates prograde waves, such that only retrograde waves can be visible above. The same is true in figure \ref{fig0}B but for the case of a negative mean flow. With a broadband spectrum of waves, whose amplitudes can vary over time due to the chaotic dynamics of convection, momentum deposition cannot be simply traced back to a handful of self-interaction terms in the Reynolds stress that would be coherent over long times. Driving of a mean flow in this context may be unexpected, but is in fact generic at sufficiently low $Pr$: as figure \ref{figPr} shows, the mean flow becomes stronger and more regular as $Pr$ decreases. This can be understood from the fact that while the forcing through wave damping is only weakly affected by decreasing $Pr$ (because waves are damped through both viscous and thermal dissipation effects), the mean flow experiences much less dissipation (it is only damped through viscosity effects), hence becoming stronger. As a result, wave-driven flows should emerge relatively easily in low-Prandtl-number fluids such as planetary cores made of liquid metal and stellar interiors \cite{Kaplan2017,King2013}, potentially affecting planetary and stellar dynamos \cite{Malyshkin2010} and magnetic reversals \cite{Carbone2006,benzi2010}.

\begin{figure*}[ht]
\centering
\includegraphics[width=15.5 cm]{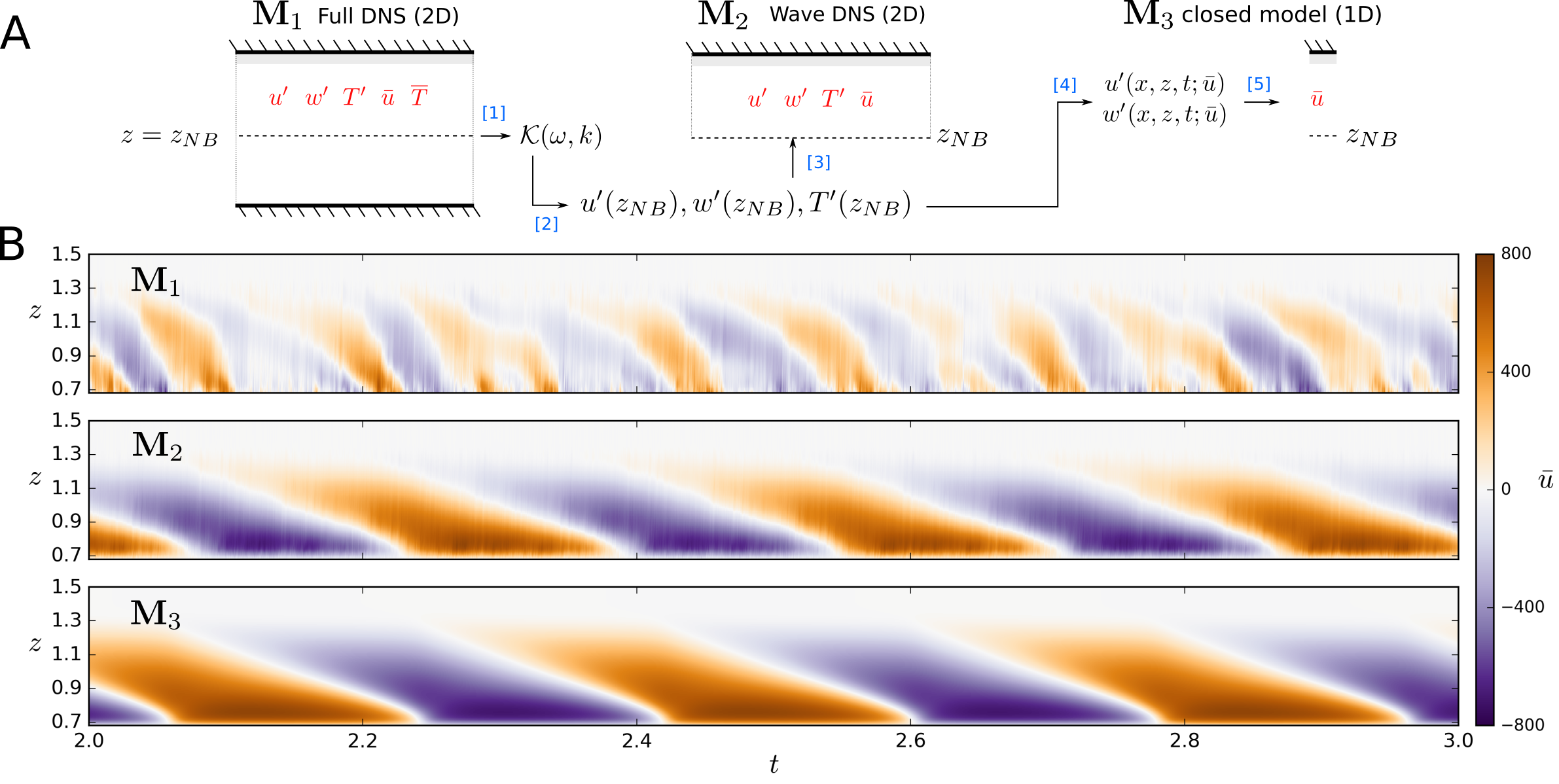}
\vspace{-0.1in}\caption{(A) Schematics of the DNS model $\mathbf{M}_1$ and the two reduced models $\mathbf{M}_2$ and $\mathbf{M}_3$. [1] We calculate the kinetic energy spectrum $\mathcal{K}$ of the fluctuations at height $z_{NB}$ obtained in $\mathbf{M}_1$. [2] The forcing ($u', w', T'$) is derived from $\mathcal{K}$ assuming that the fluctuations correspond to linear propagating internal waves. Propagation of the waves is solved [3] via DNS of the Boussinesq equations in $\mathbf{M}_2$, but is derived analytically [4] in $\mathbf{M}_3$ under WKB approximation. Thus, in $\mathbf{M}_3$, [5] we only need to solve for the mean-flow equation. As in full DNS, we use a damping layer for $1.35<z<1.5$, and boundary conditions for the mean flow are no slip. (B) $\bar{u}$ over one thermal time  obtained for each model shown in (A). Physical parameters are as in figure \ref{fig0}. Note that the colormap has been changed in figure \ref{fig1}B compared to figure \ref{fig0}C to highlight differences between $\mathbf{M}_1$ and $\mathbf{M}_{2,3}$.}\label{fig1}
\end{figure*}

We now compare results of the full DNS  model for the parameters of figure \ref{fig0} (denoted by $\mathbf{M}_1$) with results obtained from two reduced models ($\mathbf{M}_2$ and $\mathbf{M}_3$), described in figure \ref{fig1}A. The goal of the reduced models is to reproduce the evolution of the mean flow without resolving the convection. $\mathbf{M}_{2,3}$ only solve the dynamics of the stable layer and are forced by prescribing values for the flow variables at its base ($z_{NB}$). If we force with exact time-varying values of $(u', w')$ and $T'$ from the full DNS, the evolution of $\bar{u}$ is exactly reproduced in $\mathbf{M}_2$ (not shown). Observations of real systems do not generally provide information about all variables at sufficient temporal and spatial resolution over long time periods, so we only use a subset of the full DNS data to force the reduced models. Specifically, here we expand the fluctuations $u',w',T'$ in $\mathbf{M}_{2,3}$ in series of linear internal-wave modes, and we set their amplitudes such that the kinetic energy of each wave mode (defined by each wave's frequency $\omega$ and wave number $k$) matches the kinetic energy spectrum $\mathcal{K}(\omega,k)$ obtained in the full DNS at $z_{NB}$. We could have set the shape of the internal-wave spectrum by using theoretical predictions for the wave generation by turbulent convection \cite{Goldreich1990,Lecoanet2013}, but that would preclude a comparison to $\mathbf{M}_1$, whose wave spectrum differs from, e.g., \cite{Goldreich1990,Lecoanet2013}.

The reduced models only differ in how wave propagation away from the bottom boundary is solved. In $\mathbf{M}_2$, wave propagation is solved exactly by DNS of the Boussinesq equations, while in $\mathbf{M}_3$, a closed-form solution for the Reynolds stress is derived such that we only solve the 1D mean-flow \eqref{eq1}. The analytical solution for the fluctuations $(u',w')$ in $\mathbf{M}_3$ is obtained through WKB approximation, neglecting nonlinear wave-wave terms, mean-flow acceleration, and cross-interaction terms in the Reynolds stress (cf. details in Supplemental Material \cite{SM}). We note that while $\mathbf{M}_2$ is computationally cheaper than $\mathbf{M}_1$ (resolution is 8 times smaller and time steps are $\sim 3$ times larger), it remains significantly more demanding than $\mathbf{M}_3$, which is the only practical model for predicting the long-term dynamics of real systems (e.g. capturing the QBO in General Circulation Models). The goal of $\mathbf{M}_2$ is to check approximations made in $\mathbf{M}_3$.

Figure \ref{fig1}B shows the temporal variations of the mean flow $\bar{u}$ obtained in full DNS $\mathbf{M}_1$ and in the two reduced models $\mathbf{M}_2$ and $\mathbf{M}_3$. A large-scale oscillation is obtained in all three models, but the mean flow is stronger and the period is longer in the reduced models than in full DNS. Let us consider the characteristic amplitude of the mean-flow by its rms ($\bar{u}_{rms}$), and the characteristic period by taking the inverse of the peak frequency of its Fourier transform ($T_{\bar{u}}$), which we average vertically between $z=0.8$ and $1.1$. We have: $\bar{u}_{rms}=$ 179 ($\mathbf{M}_1$); 370 ($\mathbf{M}_2$); and 415 ($\mathbf{M}_3$); $T_{\bar{u}}=0.125$ ($\mathbf{M}_1$); $\approx 0.33$ ($\mathbf{M}_2$); and $\approx 0.33$ ($\mathbf{M}_3$). Clearly, even when the wave propagation is solved exactly by DNS ($\mathbf{M}_2$), the mean-flow dynamics is not reproduced quantitatively. In addition, the temporal variability  of the mean flow obtained in full DNS is lacking in both reduced models. 

The large discrepancy between the reduced models and DNS comes from the assumption that fluctuations on the lower boundary  $z=z_{NB}$ of $\mathbf{M}_{2,3}$ can be reconstructed from the time-averaged spectrum $\mathcal{K}$ using the linear wave relations for upward propagating plane waves. However, $\mathcal{K}$ can include contributions from overshooting plumes and some of the waves may be nonlinear. In our $\mathbf{M}_1$ ($\mathbf{M}_2$) simulations, the nonlinear terms have a typical magnitude of approximately 50\% (10\%) the linear terms just above $z_{NB}$, suggesting overshooting convection in $\mathbf{M}_1$ may be non-negligible at the interface. However, in the bulk of the stable region, this decreases to about 10\% (5\%), so the waves are in a weakly nonlinear regime (cf. details in Supplemental Material). Because the waves are weakly nonlinear, the energy transfer among waves does not affect the mean flow, explaining the agreement between $\mathbf{M}_2$ and $\mathbf{M}_3$. Because forcing using the spectrum higher than $z_{NB}$ could in principle attenuate contributions from nonlinear convective motions, we have run additional simulations with different forcing heights (cf. Fig. S2 of the Supplemental Material \cite{SM}): quantitative changes for the mean flow are obtained, but never lead to agreement with full DNS results. Importantly, the reconstruction of wave fluctuations from an energy spectrum neglects high-order statistics (higher than two), so  statistics in the reduced models are Gaussian. However, intermittent events exist near the interface because the convection does not have a top-down symmetry and exhibits non-Gaussian statistics. In fact, the kurtosis of the fluctuations remains large, even in the wave field far from the interface $z_{NB}$ (see Fig. 3 of the Supplemental Material \cite{SM}), suggesting that intermittency is a key component of wave generation. Intermittent intense wave events found in our DNS but neglected in the reduced models are typical of real systems. In the atmosphere, for instance, atmospheric waves sometimes propagate in the form of localized wave packets \cite{Hertzog2012}, such that wave intermittency can be non-negligible and has to be incorporated in reduced mean-flow models using stochastic processes \cite{Lott2013}.

In conclusion, the spontaneous generation and oscillation of a mean flow in our minimal, physical model is obtained for a wide range of parameters. In particular, we find that the mean flow becomes stronger as $ Pr$ decreases (figure \ref{figPr}), which highlights the necessity to account for the real value of $Pr$ in stellar and planetary dynamical models. Evaluating the impact of wave-driven flows in natural systems is challenging. Indeed, we have shown here that reduced models do not yet predict the correct physics: tackling simultaneously the three-scale dynamics due to turbulence, waves, and mean flow, seems necessary. A major source of errors in reduced models comes from the approximations made in the types of waves excited by convection, even if the stably-stratified layer is forced with waves with the same kinetic energy spectrum as in full DNS. Our analysis suggests that implementing wave intermittency (through a boundary forcing scheme that would match the high-order moments of the DNS statistics), and disantengling non-wave contributions from the source spectrum are the next step forward and will be essential to improve the long-time predictive capabilities of low-order models.

\nocite{lindzen1968,plumb1977,vallis2017,Couston2017,Burns2017,Ascher1997}

\begin{acknowledgments}
The authors acknowledge funding by the European Research Council under the European Union's Horizon 2020 research and innovation program through grant agreement No 681835-FLUDYCO-ERC-2015-CoG. DL is supported by a PCTS fellowship and a Lyman Spitzer Jr fellowship. LAC thanks Bruno Ribstein for useful discussions and references on parametrizations in General Circulation Models. Computations were conducted with support by the HPC resources of GENCI-IDRIS (Grant number A0020407543 and A0040407543) and by the NASA High End Computing (HEC) Program through the NASA Advanced Supercomputing (NAS) Division at Ames Research Center on Pleiades with allocations GID s1647 and s1439.
\end{acknowledgments}


\onecolumngrid
\appendix

\begin{center}
{\LARGE Supplementary Information}
\end{center} 
%
\section*{Simulations Details}

Table \ref{table} shows details (including typical time steps) of the simulations carried out for the full DNS model $\mathbf{M}_1$ and reduced model $\mathbf{M}_2$ with the open-source pseudo-spectral code Dedalus \cite{Burns2017}. We use Chebyshev and Fourier modes in the $z$ and $x$ direction, respectively, and a 2-step implicit/explicit Runge-Kutta scheme for time integration. The CFL condition is 0.5. For reduced model $\mathbf{M}_3$, the mean-flow equation (with radiative damping from $z=1.35$ to $z=1.5$ as in $\mathbf{M}_1$ \& $\mathbf{M}_2$) is solved using a second-order centered finite-difference scheme in $z$ and an Adams-Bashforth/Crank-Nicolson scheme for time integration. The vertical resolution is $\delta z=0.00425$ and time steps are $\delta t= 0.0002$.

To calculate the kinetic energy spectrum $\mathcal{K}$, we record the values of the fluctuations $u'$ and $w'$ in the DNS for all $x$ at a fixed interface depth $z^*$ and every $\delta t = 5\times 10^{-5}$ for one thermal time. In Fourier space, we thus have information about modes with wavenumbers ranging from $\pi$ to $128\pi$ (discarding the 0th and Nyquist modes, $L_x=2$, $n_x=256$) and positive and negative angular frequencies ranging from $2\pi$ to $10^5\times 2\pi$ in absolute value (corresponding to retrograde and prograde waves, respectively), for a total of $2.56\times 10^6$ modes. The highest-wavenumber modes put strong resolution constraints on the numerical simulation of the reduced models. Thus, we remove them from the dataset for simplicity and also because they carry so little energy that they do not affect the solution. Specifically, we remove all modes with wavenumber greater than $32\pi$. Modes with angular frequencies greater than the buoyancy frequency $N$ are not propagating modes so they are also removed from the dataset (in practice we remove all modes with angular frequency greater than $0.995N$). Finally, modes with angular frequencies lower than $0.005N$ are also removed because they carry little energy and because their treatment in $\mathbf{M}_3$ requires special care since they reach critical layers easily. In a final step, we sum up every 8 successive energy bins in the frequency direction such that the final number of modes considered decreases to $32\times 788$. Energy loss due to truncation of high-wavenumber, low-frequency and high-frequency modes when constructing the spectrum at height $z^*=0.65,0.68,0.70$ is small and reported in table \ref{table2}.

\begin{table}
\def\arraystretch{1.5}
\setlength\tabcolsep{0.1in}
\centering
\begin{tabular}{rrrrrrrr}
 Model & $Pr$		& $Ra$ 		& $S$ 	& $n_x\times n_z$ & $Q$ & $N$ & $\delta t$	\\ 
\hline	 
$\mathbf{M}_1$ & 0.1 & $19.2\times 10^7$ &  0.42 &  $512 \times 512$ & 51 & $2.0\times 10^4$ & $3\times 10^{-7}$\\
$\mathbf{M}_1$ & 0.2 & $12\times 10^7$ &  0.33 &  $256 \times 256$ & 50 & $2.0\times 10^4$ & $6\times 10^{-7}$\\
$\mathbf{M}_1$ & 0.3 & $8\times 10^7$ &  0.33 &  $256 \times 256$ & 48 & $1.9\times 10^4$ & $7\times 10^{-7}$\\
$\mathbf{M}_1$ & 0.6 & $7.2\times 10^7$ &  0.19 &  $256 \times 256$ & 51 & $2.0\times 10^4$ & $6\times 10^{-7}$\\
$\mathbf{M}_1$ & 1.0 & $5.6\times 10^7$ &  0.14 &  $256 \times 256$ & 49 & $2.0\times 10^4$ & $6\times 10^{-7}$\\
$\mathbf{M}_1$ & 3.0 & $4.4\times 10^7$ &  0.06 &  $256 \times 256$ & 50 & $2.0\times 10^4$  & $5\times 10^{-7}$\\
\hline
 $\mathbf{M}_2$ & 0.3 & - &  - &  $64 \times 128$ & - & $1.9\times 10^4$ & $2\times 10^{-6}$ \\
\hline
\vspace{-0.2in}\caption{Parameters of DNS with varying Prandtl number $Pr$ (figure 2 in the main text). $Pr$, $Ra$ and $S$ are input parameters chosen such that the $x$-averaged heat flux $Q=-\overline{T}_z+\overline{wT}$, which is depth invariant, and $x$-averaged buoyancy frequency $N$ (angular) are similar for all simulations. The domain size is fixed, i.e., $L_x\times L_z=2\times 1.5$, and the top temperature is also always $T_t=-43$. Keeping $N$ and $T_t$ constant fixes the interface depth to approximately the same value, as can be seen in supplementary figure \ref{mean}. The CFL condition is set to 0.5 ($\delta t$ gives the typical time step). For $\mathbf{M}_2$, $Pr$ and $N$ are both input parameters.}
\label{table}
\end{tabular}
\end{table}

\begin{table}
\def\arraystretch{1.5}
\setlength\tabcolsep{0.05in}
\centering
\begin{tabular}{cccccc}
full $\mathcal{K}(z=0.65)$	& truncated $\mathcal{K}(z=0.65)$ & full $\mathcal{K}(z=0.68)$	& truncated $\mathcal{K}(z=0.68)$ & full $\mathcal{K}(z=0.70)$	& truncated $\mathcal{K}(z=0.70)$ \\ 
\hline	 
$10.5\times 10^5$ & $9.9\times 10^5$ & $7.1\times 10^5$ &  $6.8\times 10^5$ & $5.8\times 10^5$ & $5.5\times 10^5$ \\
\hline
\vspace{-0.2in}\caption{Total kinetic energy spectrum $\mathcal{K}$ for the full DNS dataset and after modal truncation (see the text) at heights $z=0.65,0.68,0.70$.}
\label{table2}
\end{tabular}
\end{table}

\section*{Effect of decreasing the Prandtl number}

Figure 2 of the main text shows that the mean-flow rms becomes stronger and the flow becomes increasingly narrowbanded as the Prandtl number decreases. Because all physical effects in our model (waves, mean flow, convection) are fully coupled, changing $Pr$ changes not only the dissipation of the mean flow but also the convection and the wave generation and propagation. In order to provide a meaningful comparison of mean-flow emergence with changing $Pr$, we adapted the other input parameters ($Ra$, $S$) of our DNS in such a way that the convection and the waves are not strongly affected. Specifically, we have set $Ra$ and $S$ such that the $x$-averaged vertical heat flux $Q$ is constant, which fixes the energy available to convection, and such that the buoyancy frequency $N$ is constant, so wave propagation is not affected. All physical parameters are listed in table \ref{table}. 

\begin{figure}[H]
\centering
\includegraphics[width=0.82\textwidth]{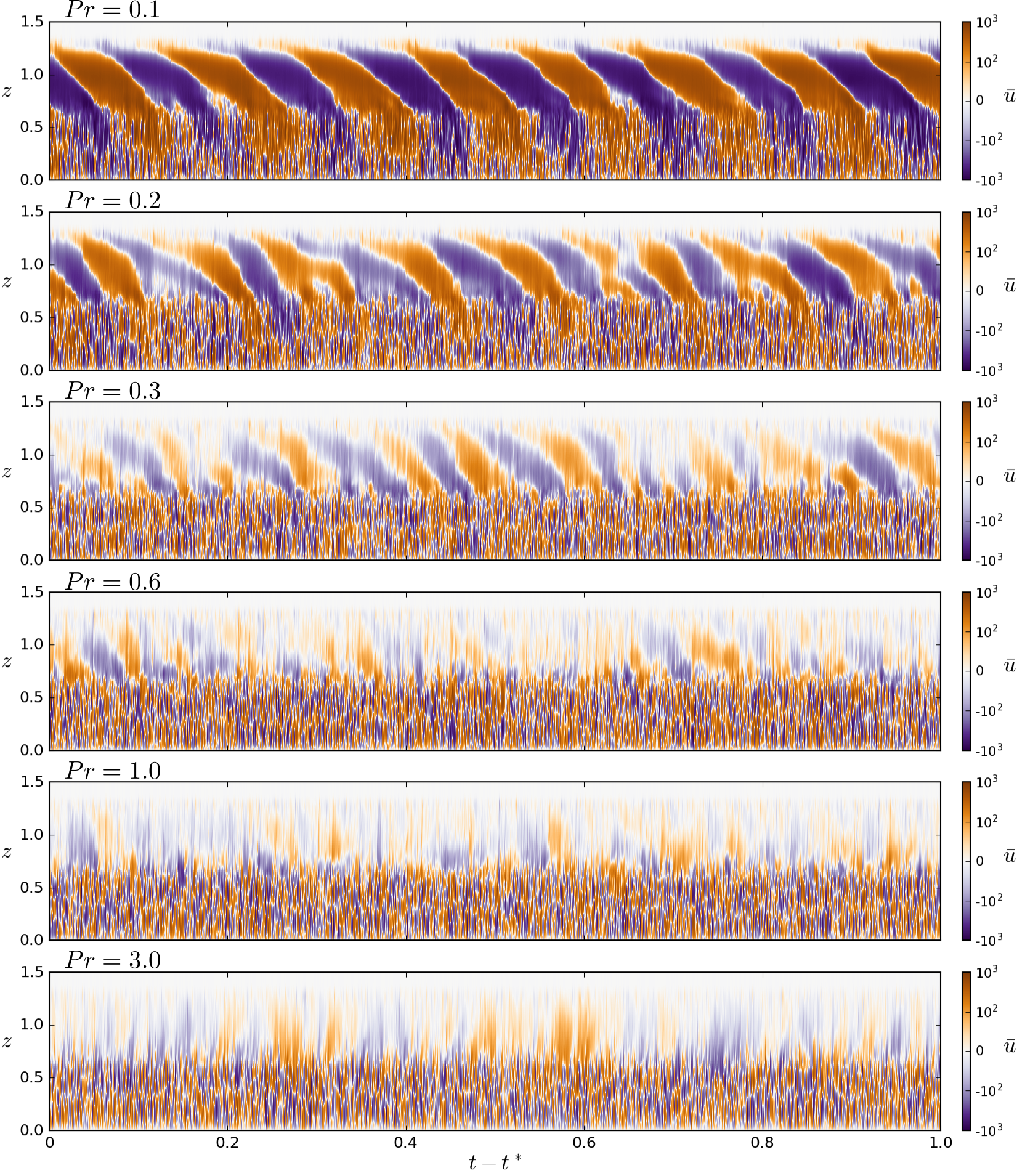}
\caption{Mean flow $\bar{u}$ obtained in DNS for different Prandtl numbers, shown as a function of $z$ over one thermal time after reaching statistically steady state. Time on the $x$ axis is the reduced time with $t^*=0.6,2.0,4.0,2.0,2.0,2.0$ for $Pr=0.1,0.2,0.3,0.6,1.0,3.0$. We use the same color axis for $\bar{u}$ in all plots in order to highlight the differences of the mean-flow amplitude in the stable region for all $Pr$.}\label{mean}
\end{figure}

The intensification of the mean flow and narrowing of the mean-flow frequency spectrum with decreasing $Pr$ can be clearly seen from the full DNS results of $\bar{u}$ shown as a function of $t$ and $z$ in supplementary figure \ref{mean}. Notably, for $Pr\geq 0.6$, the QBO-like feature of the mean flow is weak or simply non-existent, while for $Pr=0.1$ the wave-driven mean flow penetrates deep inside the convective layer. The general trend shown in supplementary figure \ref{mean} suggests that for a given planetary or stellar state, defined by its heat loss and stable stratification strength, a QBO-like behavior will be more probable at low $Pr$, which is the relevant limit for liquid iron and ionized plasmas.

%

\section*{Effect of changing the bottom boundary of the stable layer}

Let us denote $z^*$ the bottom boundary of the stable layer in the reduced models $\mathbf{M}_2$ and $\mathbf{M}_3$. In the main text, we show results for $z^*=z_{NB}$. When $z^*$ is varied (but not too different from $z_{NB}$), the mean flow exhibits similar patterns as in figure 3b $\mathbf{M}_2$-$\mathbf{M}_3$ of the main text. However, this change does have a quantitative effect, as can be seen in supplementary figure \ref{base} where we show the mean-flow rms $\bar{u}_{rms}$ and dominant period  $T_{\bar{u}}$ as functions of $z^*$. When $z^*$ increases,  there is less energy available to force the mean flow (cf. table \ref{table2}) because internal waves are damped as they propagate upward. This leads to an increase of the mean-flow period as $z^*$ increases. The increase of $T_{\bar{u}}$ when energy of a broadband internal-wave spectrum decreases is in qualitative agreement with the results for monochromatic forcing \cite{vallis2017}. The effect of $z^*$ on $\bar{u}_{rms}$ is not monotonic, which is more surprising. In simulations with less energy ($z^*$ increases), one would expect that $\bar{u}_{rms}$ decreases. However, it is clear that $\bar{u}_{rms}$ and $T_{\bar{u}}$ remain much larger for all $z^*$ than what is obtained in full DNS (shown by the dashed lines).  Thus, changing $z^*$, while having a definite effect on the results, will not lead to values significantly closer to those in the full DNS. The differences between $\mathbf{M}_2$ and $\mathbf{M}_3$ are strongest for $z=0.65$, with $\mathbf{M}_2$ closer to the full DNS.

\begin{figure}[H]
\centering
\includegraphics[width=0.8\textwidth]{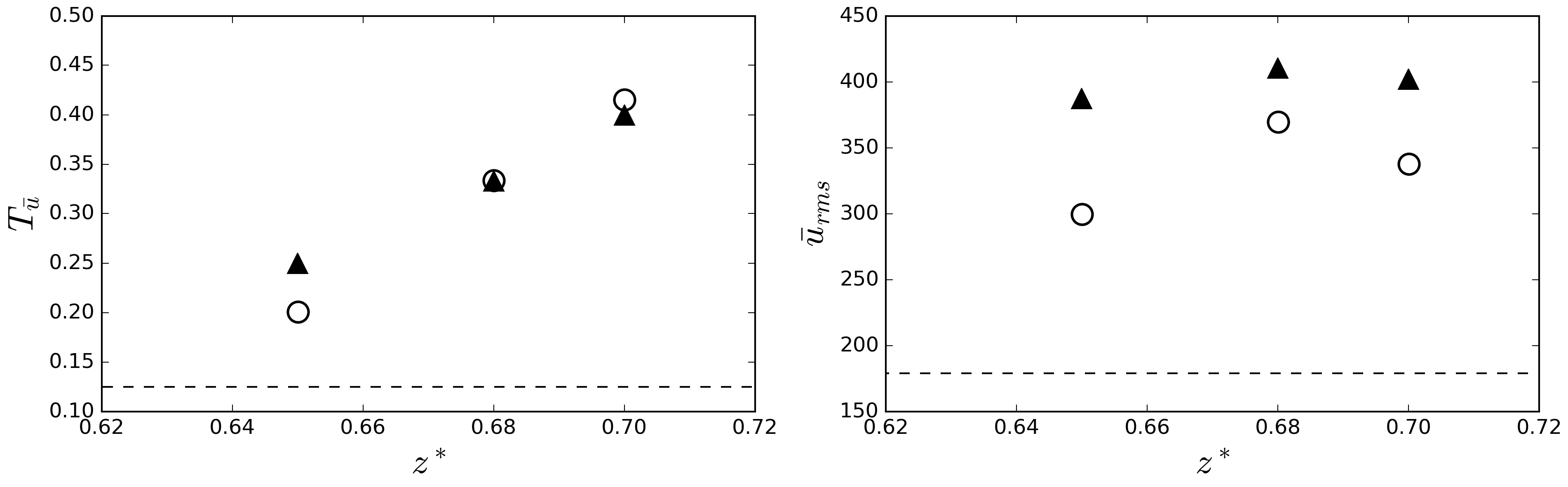}
\caption{Variations of the dominant mean-flow period (left plot) and mean-flow rms (right) averaged from $z=0.8$ to $z=1.1$,  with the height $z^*$, the bottom of the stable layer. Results are shown as open circles for $\mathbf{M}_2$ and as filled upper triangles for $\mathbf{M}_3$. The full DNS results for $\bar{u}_{rms}$ and $T_{\bar{u}}$ are shown by the dashed lines.}\label{base}
\end{figure}

\section*{Non-Gaussian Statistics in full DNS}

\begin{figure}[H]
\centering
\includegraphics[width=0.85\textwidth]{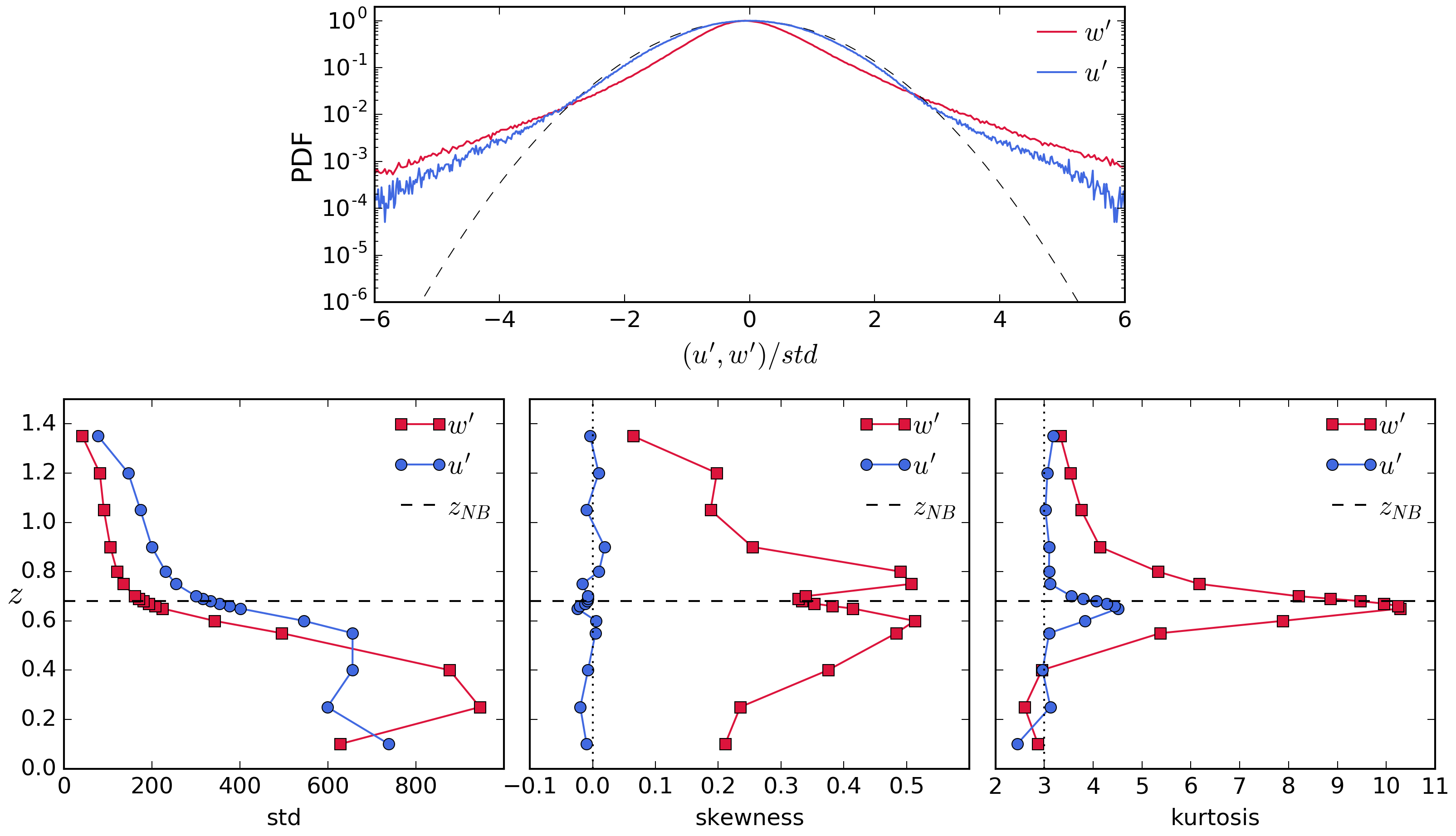}
\caption{(top) The normalized PDFs of $w'$ and $u'$ at $z=z_{NB}=0.68$ with solid lines, and the Gaussian distribution with the dashed line. The means of $w'$ and $u'$ are 0. The bottom row shows the standard deviation (left), skewness (middle) and kurtosis (right) of $w'$ and $u'$ recorded over one thermal time for the DNS results presented in the main text at 17 different depths (shown by symbols).}\label{nongaussian}
\end{figure}

In the main text we suggest that wave intermittency may explain some of the discrepancies between the full DNS and the reduced models. In the reduced models, waves are forced from the bottom boundary assuming that they have the same amplitudes at all times. Thus, waves are not intermittent in the reduced models, and the statistics of the fluctuations are Gaussian (even in the bulk of the stable layer), which means that nonlinear effects do not make the flow statistics non-Gaussian. In the full DNS, however, the PDFs of the fluctuations are non-Gaussian both in the convection and the stably-stratified layers. Supplementary figure \ref{nongaussian} shows the PDFs for $w'$ and $u'$ at $z=z_{NB}$, and the standard deviation, skewness and kurtosis as functions of depth $z$. From the top row figure, it is clear that the tails of the PDFs have much larger values than what is obtained with a Gaussian distribution (shown by the dashed line) at $z=z_{NB}$. From the bottom plots, we can see that the non-Gaussianity extends above the convective layer. While the standard deviation is much larger in the convective layer than in the stably-stratified layer (left plot), the positive skewness of vertical velocity (middle) remains relatively large in the entire domain.  Positive skewness is expected in the convection since updrafts are more vigorous than downdrafts if there is an overlying stable layer (cf. \cite{Couston2017}). The kurtosis (right plot) is even more striking---the kurtosis of $w'$ is maximum close to $z_{NB}$ and decays only slowly toward the domain boundaries, where it reaches a kurtosis of 3, the result for Gaussian statistics. In summary, the large value of the kurtosis close to the interface best demonstrates that the process of wave generation and propagation away from $z_{NB}$ is characterized by intense, rare events (shown by the heavy tails in the top figure), and that wave intermittency should be included in parameterization schemes.

\section*{Details of the reduced models}

In $\mathbf{M}_2$, we solve Eq. (2) of the main text via DNS with Dedalus for $z\in[z^*,L_z]$ assuming $T=\bar{T}+T'$, and using $N^2 = -\p_z \bar{\rho}=-PrRaS\p_z\bar{T} =4\times 10^8$, which corresponds to $-PrRaS$ times the time and vertical average (from the interface height $z^*$ to $L_z$) of the temperature gradient in $\mathbf{M}_1$. Recall that overbar (prime) denotes $x$-average quantities (fluctuations).

For the reduced model $\mathbf{M}_3$, we solve Eq. (1) of the main text from $z=z^*$ to $L_z$ for the mean flow $\bar{u}$ (which can be readily derived by taking the $x$ average of Eq. (2) of the main text), with the fluctuations $(u',w')$ derived in closed form from the linear wave equation. Eq. (1) of the main text is solved numerically using finite differences and semi-explicit/implicit time stepping.  At each timestep, $(u',w')$ are calculated from the mean flow profile, $\overline{u}(z)$, as described below; the Reynolds stresses are calculated assuming each wave only interacts with itself; then $\overline{u}(z)$ is updated by summing the effect of Reynolds stresses and viscosity.

The linearized wave equation can be written in terms of the stream function $\Psi$ ($u'=-\Psi_z$ and $w'=\Psi_x$) under the mean-field approximation \cite{plumb1977}, as
\ba{}\label{h}
\lp \p_t+\bar{u}\p_x-\DEL\rp\lb \lp \p_t + \bar{u}\p_x -  Pr \DEL \rp \DEL  - (\bar{u}_{zz}+\bar{u}_z\p_z)\p_x  \rb \Psi  
= -N^2 \p_x^2 \Psi.
\ea
As in $\mathbf{M}_2$, we take $N^2=-Pr Ra S \bar{T}_z = 4\times 10^8$. Equation \eqref{h} can be solved analytically under WKB assumption, i.e. such that $\Psi(x,z,t;\bar{u})$ is obtained in closed form for all $(x,z)$ and any mean-flow profile $\bar{u}$. Decompose the solution $\Psi=\Psi_{+}+\Psi_-$ ($\pm$ denotes waves going in the $\pm x$ direction) with $\Psi_{\pm}$ of the form 
\ba{}\label{i}
\Psi_{\pm} &= \psi_{\pm} e^{i(kx \mp \omega t)}+ *= A_{\pm} e^{\int_{z^*}^z \lp i\varphi_{\pm} + \epsilon\chi_{\pm} \rp \text{d}z'} e^{i(kx\mp\omega t)} + *
\ea
with $z^*$ the interface height (taken as $z_{NB}$ in the main text), $\epsilon$ a small parameter, $\omega>0$, and $*$ denotes the complex conjugate. Substituting \eqref{i} in \eqref{h}, and assuming that dissipation terms are order $O(\epsilon)$, $t$ derivatives of $\bar{u}$ are negligible, and $z$ derivatives of $\bar{u}$ are order $O(\epsilon)$,  yields 
\ba{}\notag
\varphi_{\pm} = \mp k \sqrt{\f{1-\varpi_{\pm}^2}{\varpi_{\pm}^2}},~~
\epsilon\chi_{\pm} =  -\f{1}{2}\f{d}{dz}\ln|\varphi_{\pm}| - \f{k^3(1+\Pr)}{2N\sqrt{1-\varpi_{\pm}^2}\varpi_{\pm}^4},
\ea
for upward-propagating waves, with $\varpi_{\pm}=\omega/N\mp k\bar{u}/N$ the normalized Doppler-shifted angular frequency. 

Convection generates a broad range of internal wave modes with different wavenumbers and frequencies such that the global streamfunction solution can be written as
\ba{}\label{totpsi}
\Psi = \sum_{k,\omega} \lb \psi_+ e^{i(kx - \omega t)} + \psi_- e^{i(kx + \omega t)} + * \rb.
\ea
We decompose the Reynolds stress forcing of the mean flow as
\ba{}\label{eq3}
-\p_z \overline{(w'u')} = -\p_z \lp \mathcal{F}_+ + \mathcal{F}_- \rp -\p_z \mathcal{C}, 
\ea
where 
\ba{}\label{eq4}
\mathcal{F}_{\pm} = -\sum_{k,\omega} ik \psi_{\pm} \p_z \psi^{*}_{\pm} + * = \sum_{k,\omega} f_{\pm}(z)
\ea
is the momentum fluxes due to the self interactions of prograde/retrograde waves and $\mathcal{C}$ is the momentum flux of all cross (i.e. non-self) interaction terms. Model $\mathbf{M}_3$  neglects $\mathcal{C}$ and approximates the individual momentum fluxes using the WKB solution as 
\ba{}\label{eq5}
f_{\pm} = -k{\varphi}_{\pm}(z=0)|A_{\pm}|^2 exp\lb  -\int_{z^*}^{L_z}  \f{k^3(1+Pr) dz'}{N\varpi_{\pm}^4\sqrt{1-\varpi_{\pm}^2}} \rb.
\ea
Note that neglecting cross-interaction terms from the Reynolds stress is a customary approximation in GCMs, which makes the wave-driven forcing depends on time  through $\bar{u}$ only.

Internal waves have critical layers when their Doppler-shifted frequency goes to zero.  In $\mathbf{M}_1$-$\mathbf{M}_2$, critical layers are regularized by dissipation or nonlinearity. In $\mathbf{M}_3$, when an internal wave approaches a critical layer, all of its remaining momentum is deposited following the work of \cite{lindzen1968}. We find that, in our configuration, setting the force due to internal waves at critical layers to zero yields almost the same results as depositing all of the momentum. This suggests that momentum deposition at critical layers is negligible compared to momentum deposition by viscous dissipation in the WKB model.

\section*{Details of the forcing and boundary conditions in the reduced models}

The bottom forcing in the reduced models is constructed from the kinetic energy spectrum at the interface height $z^*$ extracted from full DNS, assuming that all motions correspond to linear internal waves (recall, again, that we take $z^*=z_{NB}$ for the results of the main text). That is, we assume that the fluctuations can be written as a sum of wave modes of the form (sum runs over both prograde and retrograde modes)
\ba{}\label{forcing}
(u',w',T') = \sum_{k,\omega} (U_{\pm},W_{\pm},T_{\pm}) e^{i(kx\mp\omega t+\gamma_{\pm}z)} + *
\ea
where $\gamma_{\pm}$ is the vertical wavenumber.  We use the polarization relations of linear internal waves to express $(U_{\pm},W_{\pm},T_{\pm})$ at $z^*$ in terms of $\mathcal{K}$. $\mathcal{K}$ reads (hat denotes $x,t$ Fourier transform) 
\ba{}
\mathcal{K}(\omega,k)=\f{1}{2}\lp |\hat{u'}(z=z^*)|^2+|\hat{w'}(z=z^*)|^2 \rp.
\ea
Using the linear dispersion relation $\gamma_{\pm} = \mp k \sqrt{(N^2-\omega^2)/\omega^2}$ we obtain for the streamfunction amplitude (cf. \eqref{i}) and the other variables ($\zeta_{\pm}$ an arbitrary phase)
\bsa{}\notag
&A_{\pm} = \f{\omega e^{i\zeta_{\pm}}}{kN}\sqrt{\f{\mathcal{K}}{2}},~U_{\pm} = -i\gamma_{\pm}A_{\pm}, \\ 
&W_{\pm} = ikA_{\pm}, ~T_{\pm} = \f{\pm k\bar{T}_z}{\omega}A_{\pm}.
\esa

Vertical boundary conditions are no-slip for the fluctuations at $z=L_z$ in $\mathbf{M}_2$, and $\bar{u}=0$ at $z=z^*,L_z$ in $\mathbf{M}_2$, $\mathbf{M}_3$. The no-slip bottom boundary condition for $\bar{u}$ introduces  errors, as $\bar{u}\neq 0 $ at $z=z^*$ in full DNS. However, using other conditions does not lead to improvements of $\mathbf{M}_2$-$\mathbf{M}_3$ results or introduces new tuning parameters. A stress-free boundary condition leads to unphysical jets forming close to the bottom boundary. A small damping layer (with, e.g., radiative damping coefficient $D$) with a free-slip boundary condition could potentially exert a sufficient drag on $\bar{u}$ such that it reaches a physically-sound dynamical equilibrium, but it introduces the parameter $D$, which, with no prior study of its effect, could only be set in an ad-hoc manner.

\section*{Quantification of Nonlinear effects in the wave region}

The mean flow obtained in the different models may differ in part because of nonlinear effects in the wave region $z\geq 0.7$. Therefore, in table \ref{table3}, we quantify the importance of nonlinear effects through the ratio $\mathcal{N}''$  of the rms of the wave-wave to wave term $|(\b{u'}\cdot\NA)\b{u'}-\overline{(\b{u'}\cdot\NA)\b{u'}}|$ (overline denotes $x$ average) divided by the rms of one of the dominant non-static force in the momentum equation, i.e. the dynamic pressure term $|\NA p' |$. The wave-mean to wave and wave-wave to mean nonlinear ratios $\mathcal{N}'$ and $\overline{\mathcal{N}''}$, given by the rms of $|(\bar{\b{u}}\cdot\NA)\b{u'}+(\b{u'}\cdot\NA)\bar{\b{u}}|$ and  $|\overline{(\b{u'}\cdot\NA)\b{u'}}|$  divided by the rms of $|\NA p' |$, are also reported in table \ref{table3} for completeness (other nonlinearities in the momentum equation are zero). Overall, $\mathcal{N}''$, $\mathcal{N}'$, and $\overline{\mathcal{N}}$ decrease with height for $z\geq 0.7$, and we provide values in table \ref{table3} at three different heights: close to the interface at $z=0.7$, slightly above at $z=0.8$, and in the bulk of the wave region at $z=1.0$. For $Pr=0.1$, $\bar{u}$ is strong, and we find $\mathcal{N}'>\mathcal{N}''>\overline{\mathcal{N}''}$ at $z=0.8,1.0$. For cases of $Pr\geq 0.2$, $\bar{u}$ is weaker than for $Pr=0.1$ and the wave-wave to wave nonlinearity dominates; also, the relative importance of $\overline{\mathcal{N}''}$ compared to $\mathcal{N}'$ increases as $Pr$ increases because the mean-flow weakens. For all $\mathbf{M}_1$ simulations, $\mathcal{N}''$ is relatively large (typically 50\%) close to the interface, suggesting that wave-wave to wave nonlinear effects may be non-negligible; higher up, however, $\mathcal{N}''$ is much smaller, typically 10\% at $z=1.0$, such that $\mathcal{N}''$ is large at $z=0.7$ mostly because there are traces of (strongly-nonlinear) convective motions at the base of the stable layer, and the waves above the interface are actually weakly nonlinear in all our simulations. We find that $\mathcal{N}''$ is smaller in $\mathbf{M}_2$ than in $\mathbf{M}_1$ ($Pr=0.2$), even though the kinetic energy is the same at the interface. This finding is in agreement with the idea that nonlinear effects in $\mathbf{M}_1$ are due to overshooting plumes at $z=0.7$, and that waves have typically larger amplitudes (and more nonlinearity) in $\mathbf{M}_1$ than in $\mathbf{M}_2$ because they propagate in the form of intermittent intense wave packets  in $\mathbf{M}_1$. $\mathcal{N}'$ ($\overline{\mathcal{N}''}$) is larger (smaller) in $\mathbf{M}_2$ than in $\mathbf{M}_1$, because $\bar{u}$ is stronger but reverses more slowly in $\mathbf{M}_2$ than in $\mathbf{M}_1$.

\begin{table}[H]
\def\arraystretch{1.5}
\setlength\tabcolsep{0.1in}
\centering
\begin{tabular}{rcccccc|c}
 Model & $\mathbf{M}_1$ & $\mathbf{M}_1$ & $\mathbf{M}_1$ &$\mathbf{M}_1$ & $\mathbf{M}_1$ &$\mathbf{M}_1$ &$\mathbf{M}_2$ \\ \hline
 $Pr$ & 0.1 & 0.2 & 0.3 & 0.6 & 1.0 & 3.0  & 0.2 \\	
 $\mathcal{N}''(\mathcal{N}')(\overline{\mathcal{N}''})$, z=0.7 & 52(35)(10) & 51(17)(11) & 27(8)(6) & 60(12)(13) & 58(8)(14) & 61(10)(20)	&	9(17)(2) \\
 $\mathcal{N}''(\mathcal{N}')(\overline{\mathcal{N}''})$, z=0.8 & 26(30)(5) & 20(8)(4) & 17(4)(4) & 20(4)(5) & 19(3)(5) & 19(4)(6) & 5(17)(1)	\\ 
 $\mathcal{N}''(\mathcal{N}')(\overline{\mathcal{N}''})$, z=1.0 & 17(28)(3) & 12(6)(3) & 11(2)(3) & 12(2)(3) & 11(1)(3) & 9(1)(3) & 5(8)(2)	\\ 
\hline
\vspace{-0.2in}\caption{Quantification of nonlinear effects in $\mathbf{M}_1$ and $\mathbf{M}_2$ simulations. All numbers are in \%. See the text for the definition of $\mathcal{N}''$, $\mathcal{N}'$, $\overline{\mathcal{N}''}$.}
\label{table3}
\end{tabular}
\end{table}



\bibliography{Main_text_incl_figures.bbl}

\end{document}